\documentstyle[twoside,fleqn,espcrc2,epsf]{article}

\newcommand{\AmS}{{\protect\the\textfont2
  A\kern-.1667em\lower.5ex\hbox{M}\kern-.125emS}}
\newcommand{\Tr}{{\rm Tr}}

\def\lsim{\raise0.3ex\hbox{$<$\kern-0.75em\raise-1.1ex\hbox{$\sim$}}}
\def\gsim{\raise0.3ex\hbox{$>$\kern-0.75em\raise-1.1ex\hbox{$\sim$}}}

\title{\vspace{-3.6cm}
\begin{flushright}
{\normalsize
UTCCP-P-12\\
\vspace{-.2cm}
August 1996}
\end{flushright}
\vspace{1.5cm}
Four-dimensional Simulation of the Hot Electroweak Phase Transition
with the SU(2) Gauge-Higgs Model
\thanks{Talk given at Lattice 96, St.~Louis, USA.}}

\author{Yasumichi Aoki\address{Center for Computational Physics,
        University of Tsukuba, Ibaraki 305, Japan}}
       
\begin{document}

\begin{abstract}
We study the finite-temperature phase transition of the four-dimensional
SU(2) gauge-Higgs model for intermediate values of the Higgs boson mass in 
the range $50\lsim m_H \lsim 100$GeV on a lattice with the temporal lattice 
size $N_t=2$.  The order of the transition is 
systematically examined using finite size scaling methods. Behavior 
of the interface tension and the latent heat for an increasing Higgs 
boson mass is also investigated.
\end{abstract}

\maketitle

\section{INTRODUCTION}

The possibility that the baryon number asymmetry of the Universe has 
been generated 
in the course of the electroweak phase transition has led to recent lattice 
investigations of the SU(2) gauge-Higgs model by several
groups\cite{ShReview}. Their studies have shown that a first-order phase
transition takes place at a finite temperature in this system for a light
Higgs boson,  which, however, becomes rapidly weak  as the Higgs boson
mass increases.
A central question, relevant for the baryon number asymmetry problem, 
is whether the first-order transition survives with a sufficient strength 
for realistically large Higgs boson mass, experimentally bounded by $M_H\geq
64$GeV\cite{Janot}. To answer this question many studies have been done within
the  perturbatively reduced three-dimensional model\cite{ShReview}.  On the
other hand, only a few  studies with the original four-dimensional model exist
in this region of Higgs boson mass\cite{ShReview}.
In this article, we report results of our simulation of the four-dimensional 
model, aiming at a systematic finite-size scaling analysis of the 
order of the transition for the Higgs boson mass in 
the range $50\lsim m_H \lsim 100$GeV.

\section{SIMULATION}

We employ the standard action given by 
\begin{eqnarray}
\label{S_lat}
  S & = & \mathop{\sum}_n \{
    \mathop{\sum}_{\mu > \nu} \frac{\beta}{2} \Tr U_{n,\mu \nu} 
    \mbox{} + \mathop{\sum}_\mu \kappa \Tr
      ( \Phi_n^{\dag} U_{n, \mu} \Phi_{n+\hat{\mu}} ) \nonumber \\
    && \mbox{} - \rho_n^2 - \lambda ( \rho_n^2 -1 )^2
    \},
\end{eqnarray}
with the complex $2\times2$ matrix $\Phi$ decomposed as
$\Phi_n = \rho_n \alpha_n , \; \rho_n \geq 0; \; \alpha_n \in$ SU(2).
All of our simulations are made for the temporal extent $N_t=2$.
We set the gauge coupling $\beta = 1/g^2 = 8$, and make simulations for 
6 values of the scalar self-coupling $\lambda$ as listed in
Table \ref{tab:lambda}.
Also listed in the table are the zero-temperature Higgs boson mass $M_H$ at 
the transition point of $N_t=2$ lattice, estimated by interpolating 
available data for $M_H$\cite{Leipzig92,DESY95,DESY96} as a function of 
$\lambda$.
For each value of $\lambda$ runs are made on an $N_s^3 \times 2$ lattice with 
$N_s=8,12,16,24,32$,  and in addition with $N_s=40$ for $\lambda=0.001,
0.0017235, 0.003 ( M_H = 67, 85, 102$GeV ). Gauge and scalar fields are updated
with a combination of the heat  bath\cite{BunkLat95} and overrelaxation\cite{OR}
algorithms in the ratio  reported to be the fastest in ref.~\cite{DESY95}.
For each parameter point we make $10^5$ iterations of the combined updates.

\begin{table*}[t]
\setlength{\tabcolsep}{1.2pc}
\newlength{\digitwidth} \settowidth{\digitwidth}{\rm 0}
\catcode`?=\active \def?{\kern\digitwidth}
\caption{Choice of scalar self coupling
  and corresponding zero temperature Higgs boson mass.}
\label{tab:lambda}
\begin{tabular*}{\textwidth}{lcccccc}
\hline
\hline
$\lambda$ & $0.0005$ & $0.000625$ & $0.00075$ & $0.001$ & $0.0017235$ & $0.003$ \\
\hline
$M_H$(GeV) & $47$   &   $53$     &   $58$    &   $67$  &    $85$     &  $102$\\
\hline
\hline
\end{tabular*}
\vspace{-12pt}
\end{table*}

\section{FINITE-SIZE SCALING ANALYSIS}

Let us define the angular part of the spatial component of the hopping term
in the action by
\begin{equation}
  \label{Lambda}
  \Lambda_s \equiv
    \frac{1}{3 N_s^3 N_t} \mathop{\sum}_n \mathop{\sum}_{j=1}^{3} \left[
    \frac{1}{2} \Tr ( \alpha_n^{\dag} U_{n, j} \alpha_{n+\hat{j}} ) \right].
\end{equation}

\begin{figure}[t]
  \hspace{-6pt}
  \epsfxsize=7.4cm \epsfbox{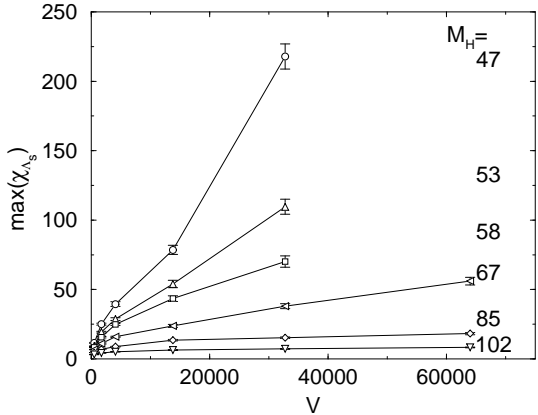}
  \vspace{-30pt}
  \caption{Maximum height of the susceptibility of $\Lambda_s$ as a function
    of the volume. Lines are guides for eyes.}
  \label{fig:sus}
  \vspace{-12pt}
\end{figure}
\begin{figure}[t]
  \hspace{-6pt}
  \epsfxsize=7.4cm \epsfbox{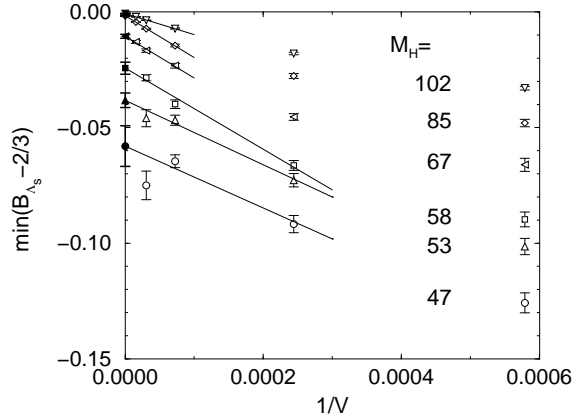}
  \vspace{-30pt}
  \caption{Valley depth of Binder cumulant of $\Lambda_s$ as a function
    of the inverse volume. Filled symbols show the infinite volume limit.
    }
  \label{fig:Binder}
  \vspace{-12pt}
\end{figure}

\noindent In Fig.~\ref{fig:sus} we show the volume dependence of the maximum
height of the susceptibility defined by
\begin{equation}
  \label{chi}
    \chi_{\Lambda_s} \equiv V \left( \langle \Lambda_s^2 \rangle -
    \langle \Lambda_s \rangle^2 \right),
\end{equation}
with $V=N_s^3$, which is calculated by the standard reweighting
technique.
For $47 \leq M_H \leq 67$GeV the maximum value increases linearly toward
larger volumes, which is expected for the case of a first-order transition. 
In contrast, we observe a very flat volume dependence for 
$85\leq M_H \leq 102$GeV, albeit the maximum value is increasing slowly in the
range of volume used here.

In Fig.~\ref{fig:Binder} we plot the valley depth
of the Binder cumulant defined by
\begin{equation}
  \label{Binder}
  B_{\Lambda_s} \equiv 1 - \frac{1}{3} \frac{\langle \Lambda_s^4 \rangle}
  {\langle \Lambda_s^2 \rangle^2}.
\end{equation}
as a function of inverse volume.
Lines are linear fits to the largest 3 volumes for each $M_H$.
For $47 \leq M_H \leq 67$GeV the value extrapolated to the infinite volume 
clearly deviates from $2/3$, providing additional evidence for a first-order
transition. For $85 \leq M_H \leq 102$GeV the deviation
decreases by an order of magnitude, although still finite within the error.

These results clearly show that the transition is of first order for 
$47 \leq M_H \leq 67$GeV. It is also clear that the transition, if first
order,  is a very weak one at
$M_H=85$GeV and 102GeV. It is quite possible that the transition has turned
into  a crossover for this range of $M_H$.  Data for larger
volumes are needed, however, for a conclusive analysis on this point.

\section{LATENT HEAT AND INTERFACE TENSION}

\begin{figure}[t]
  \hspace{-6pt}
  \epsfxsize=7.4cm \epsfbox{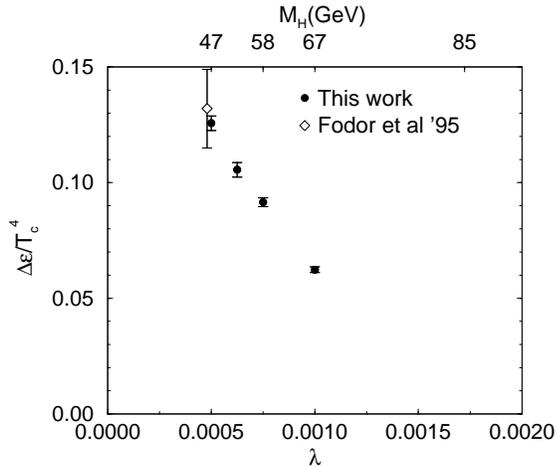}
  \vspace{-30pt}
  \caption{Latent heat as a function of Higgs boson mass.
    Open symbol is from
    ref.~[4] and is shifted
    slightly left for visualization.}
  \label{fig:d_e}
  \vspace{-12pt}
\end{figure}

The Higgs boson mass dependence of the latent heat  $\Delta \epsilon$ provides a
physical measure of the weakening of the first-order transition as $M_H$
increases. 
Here we calculate $\Delta \epsilon$ for
$M_H = 47, 53, 58, 67$GeV using the Clausius-Clapeyron
equation \cite{d_e_d_rho},
\begin{equation}
  \label{CC}
  \Delta \epsilon \simeq - M_H^2 \kappa \Delta \langle \rho^2 \rangle.
\end{equation}
where the gap $\Delta \langle \rho^2 \rangle$ is estimated from the run on  
the largest volume for each $M_H$. The result is shown in Fig.~\ref{fig:d_e}.

\begin{figure}[t]
  \hspace{-6pt}
  \epsfxsize=7.4cm \epsfbox{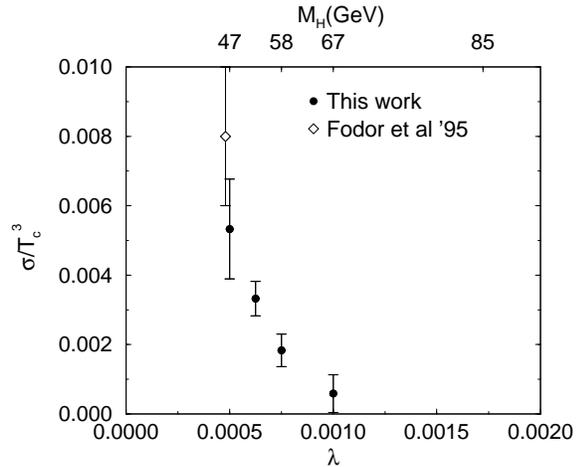}
  \vspace{-30pt}
  \caption{Interface tension as a function of Higgs boson mass.
    Legends are the same as in
    Fig.~3.}
  \label{fig:ift}
  \vspace{-12pt}
\end{figure}

Interface tension provides another indicator of the strength of the
first-order transition.  Let $P_{max}$ and $P_{min}$ be the peak and
valley height of the distribution of $\Lambda_s$, reweighted such that
the two peaks have an equal height. Define 
$\hat\sigma_V \equiv - ({N_t^2}/{2 N_s^2}) \ln ({P_{min}}/{P_{max}}) $. 
For  spatially cubic lattices, finite-size formula for the true
interface tension $\sigma$ is
given by\cite{Kanaya}
\begin{equation}
  \label{ift_corr}
  \hat\sigma_V (N_s, N_t) = \frac{\sigma}{T_c^3} - \frac{N_t^2}{N_s^2}
  \left[ c - \frac{1}{4} \ln N_s + \frac{1}{2} \ln 3 \right],
\end{equation}
where $c$ is a constant independent of $N_s$. Making a two parameter fit of
$\hat\sigma_V$ obtained for the largest three volumes for $M_H = 47, 53, 
58$GeV or two volumes for $M_H = 67$GeV we find $\sigma/T_c^3$
shown in Fig.~\ref{fig:ift}. 

Both the latent heat and the interface tension rapidly
decrease with an increasing Higgs boson mass and seem to vanish around 
$M_H \sim 80$GeV.

\section{SUMMARY}

Our finite-size scaling study establishes a first-order transition for
$47 \leq M_H \leq 67$GeV. For larger Higgs boson masses a rapid weakening of the
transition makes it difficult to draw a definitive conclusion on the order
within the range
of lattice volumes employed in our simulation.  However, combining
finite-size data  with results for the latent heat and the interface tension,
our four-dimensional study suggests that the first-order transition terminates
around $M_H \sim 80$GeV in the $N_t=2$ SU(2) gauge-Higgs model. This is
consistent with the result of a recent finite-size scaling study carried
out in the dimensionally reduced three-dimensional model\cite{Kajantie96}.   

\section*{ACKNOWLEDGEMENTS}
I would like to thank Akira Ukawa for useful discussions.
The numerical calculations were carried out on VPP/500
at Science Information Processing Center of University of Tsukuba
and at Center for Computational Physics at University of Tsukuba.

\end{document}